**Observation of Superconductivity in a Cu-Rich High-Entropy Alloy**

*Arumugam Sonachalam**

Centre for High Pressure Research, School of Physics, Bharathidasan University, Tiruchirappalli 620024, India.

Department of Physics, Gandhigram Rural Institute, Deemed to Be University, Gandhigram-624302, Dindigul District, Tamil Nadu, India.

Department of Bioelectronics and Biosensors, Alagappa University, Karaikudi, Tamil Nadu, 630003, India

*Corresponding author: sarumugam1963@yahoo.com

**Abstract:**

A Cu-rich multielement alloy containing Zn, Sn, Ag, and Au was synthesized and studied to explore its structural and physical properties. The alloy was prepared by vacuum-sealed melting of high-purity elements followed by rapid quenching in cold water. Powder X-ray diffraction confirms that the alloy crystallizes in a face-centered cubic (*fcc*) Cu-type structure. Field-emission scanning electron microscopy and energy-dispersive X-ray spectroscopy (EDS) mapping show uniform distribution of all constituent elements. Magnetic measurements indicate the appearance of a superconducting transition at low temperature, while electrical resistivity remains metallic without reaching zero resistance. This difference between magnetic and transport behaviour suggests that the superconducting state may be limited to local or partial regions within the sample. These findings indicate that the Cu-rich multielement alloy exhibits unusual low-temperature properties and may represent a new class of compositionally complex Cu-based materials with superconducting characteristics.

## 1. Introduction

High-entropy alloys (HEAs) are a class of metallic materials in which several principal elements are combined in nearly equiatomic or significant proportions [1-5]. Their high configurational entropy stabilizes simple solid-solution phases, such as body-centered cubic (bcc) or face-centered cubic (*fcc*), instead of complex intermetallic compounds [6-22]. The strong chemical disorder in these systems introduces unique effects—high-entropy stabilization, lattice distortion, sluggish diffusion, and the so-called cocktail effect—that often lead to unusual structural, mechanical, and functional behaviour [3].

The discovery of superconductivity in the equiatomic Ta–Nb–Hf–Zr–Ti HEA [1-3] demonstrated that superconductivity can exist even in highly disordered metallic systems. Since then, several HEA and medium-entropy alloy (MEA) superconductors with different crystal structures and elemental combinations have been reported [15,16,27-29]. Despite the random atomic distribution, many of these materials exhibit robust superconducting transitions, indicating that electron pairing can persist even under strong structural and chemical disorder. Moreover, several studies have revealed a close correlation between the superconducting transition temperature ($T_c$), the valence electron concentration (VEC), and local lattice distortion, providing valuable guidance for designing new superconducting alloys [6,8,29,30]. Recent investigations under pressure have further revealed that HEA superconductivity remains stable over a wide range of conditions, highlighting the intrinsic resilience of these multielement systems [2, 23, 31].

However, most of the reported HEA superconductors are based on early transition metals such as Ta, Nb, Hf, Zr, and Ti. Very little work has been done on alloys that are rich in late transition metals such as copper. This is particularly interesting because copper plays a central role in high-temperature superconductors like the cuprates, where superconductivity arises in Cu-O planes with strong electron correlations [32-34]. These copper-oxide compounds have shown that modifying the local electronic environment around Cu atoms can lead to dramatic changes in electronic properties. Although metallic copper itself is not superconducting, alloying it with other elements can alter its electronic density of states and lattice vibrations, possibly creating conditions favourable for new superconducting behaviour.

In the present study, we have synthesized a Cu-rich multielement alloy containing Zn, Sn, Ag, and Au to explore its structural and physical properties. Powder X-ray diffraction (XRD) confirms that the alloy crystallizes in a face-centered cubic (*fcc*) Cu-type structure. Microstructural studies using field-emission scanning electron microscopy (FESEM) and energy-dispersive X-ray spectroscopy (EDS) mapping show a nearly uniform distribution of all elements. Magnetic measurements indicate a low-temperature superconducting transition, whereas the electrical resistivity remains metallic down to 2 K. These findings suggest that the superconducting regions may be locally distributed within the Cu-rich matrix. The results open a new direction for studying compositionally complex Cu-based alloys and their connection to the broader family of Cu-related superconductors.

## 2. Experimental Techniques

High-purity elemental Cu (99.99%), Zn (99.99%), Sn (99.99%), Ag (99.99%), and Au (99.99%) were weighed in nominal proportions of Cu ≈ 80 wt%, Zn ≈ 20 wt%, Sn ≈ 4 wt%, Ag ≈ 8 wt%, and Au ≈ 2 wt%. The mixture was thoroughly ground, pelletized, and sealed in an evacuated quartz tube (~$10^{-4}$ mbar). The sealed tube was then heated until complete melting occurred and subsequently quenched rapidly in cold water. The resulting ingot was sectioned and polished for further characterization. Ambient X-ray powder diffraction (XRPD) measurements were performed at the XPress beamline, Elettra Sincrotrone Trieste, Italy [35].

Finely ground powders were obtained by crushing small pieces of the alloy using an agate mortar and pestle. The powders were packed tightly into a glass capillary (inner diameter ≈ 200 μm) for diffraction data collection. Scanning electron microscopy (SEM) and energy-dispersive X-ray spectroscopy (EDX) analyses were carried out using a ZEISS Supra 40 field-emission scanning electron microscope equipped with an EDX detector to verify surface morphology and elemental composition. Electrical resistivity (ρ(T)) measurements were performed in the temperature range 300–2 K using a 9 T closed-cycle refrigerator variable temperature insert (CCR-VTI, Cryogenic Ltd., UK) system. A standard four-probe configuration was employed, with electrical contacts made using high-purity silver paste (4929N) and copper wires (diameter ≈ 0.05 mm) on a rectangular-shaped polished sample. The contact resistance was minimized by cleaning the surface and curing the silver paste before cooling. Magnetization (M–T) measurements were conducted using a Physical Property Measurement System (PPMS, Quantum Design) to investigate the low-temperature magnetic response.

## 3. Results and Discussion:

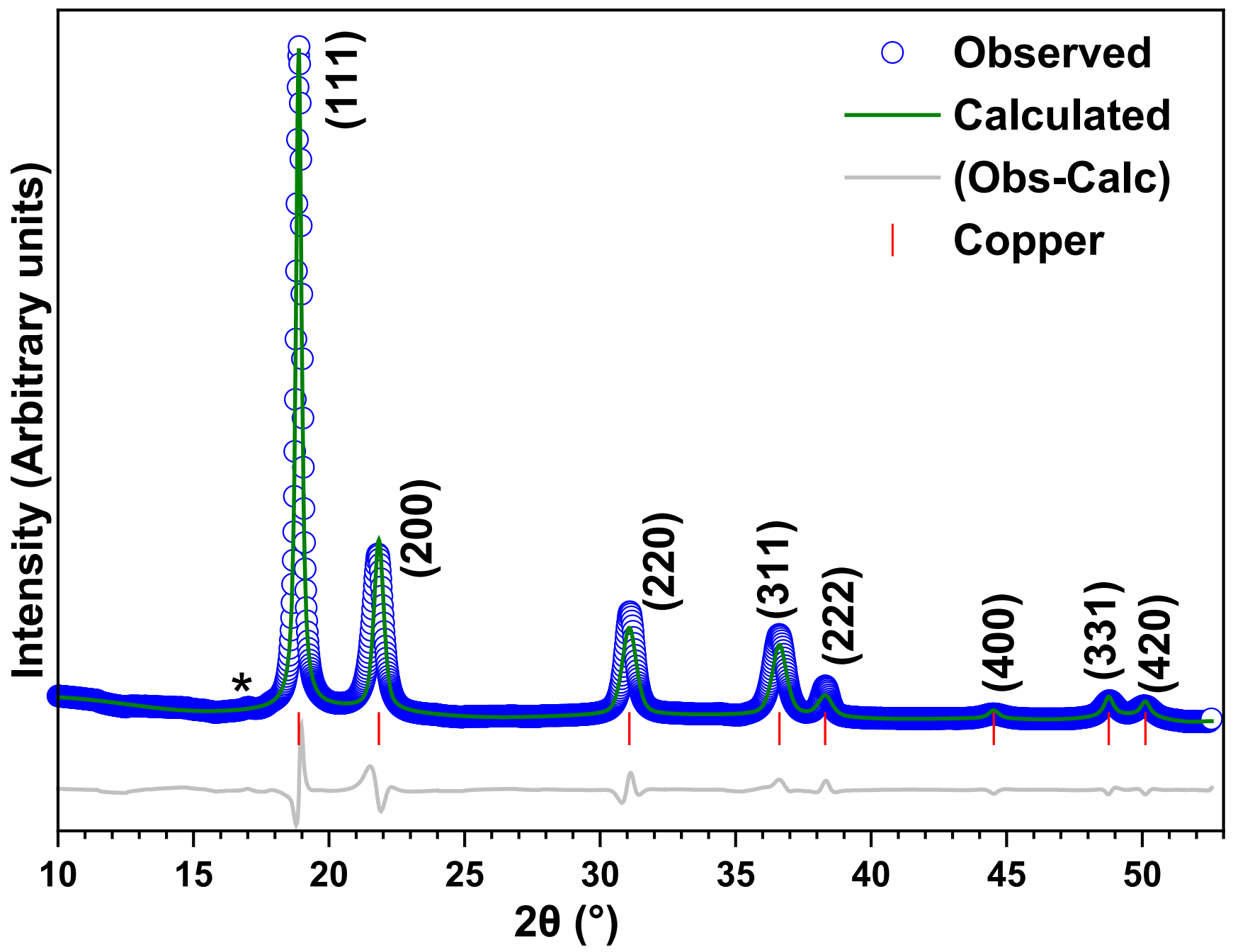

Figure 1: Measured X-ray diffraction (XRD) pattern of the Cu–Zn–Sn–Au–Ag alloy recorded under ambient conditions (symbols), along with the corresponding Rietveld refinement profile (solid green line). The analysis was performed using a wavelength of 0.4957 Å. The red vertical bars below the diffractogram indicate the Bragg peak positions of the fitted phase.

Figure 1 presents the measured X-ray powder diffraction (XRPD) pattern of the Cu-Zn-Sn-Ag-Au alloy at ambient conditions together with the corresponding Rietveld refinement results. The refinement was performed using the crystallographic information file (CIF) obtained from the ICSD database (collection code 7954) as the starting model. The analysis indicates that the alloy adopts a face-centered cubic (*fcc*) structure with space group Fm-3m (225) and a refined lattice parameter $a$ =3.6955 Å and corresponding to a unit cell volume of V= 50.467 $Å^3$. The XRPD patterns confirm that the alloy crystallizes as a single-phase *fcc* Cu-type structure without any detectable secondary phases. The refined lattice parameters are close to those of pure Cu, suggesting that Zn, Sn, Ag, and Au atoms are substitutionally incorporated into the Cu lattice. Field-emission scanning electron microscopy (FESEM) images reveal a fine-grained microstructure are shown in Figure 2. Elemental mapping using energy-dispersive X-ray spectroscopy (EDS) confirms the presence and homogeneous distribution of all constituent elements (Cu, Zn, Sn, Ag, Au), with only minor compositional variations observed near grain boundaries. Such nanoscale heterogeneity may locally modify the electronic density of states and lattice strain, potentially influencing the observed low-temperature behaviour.

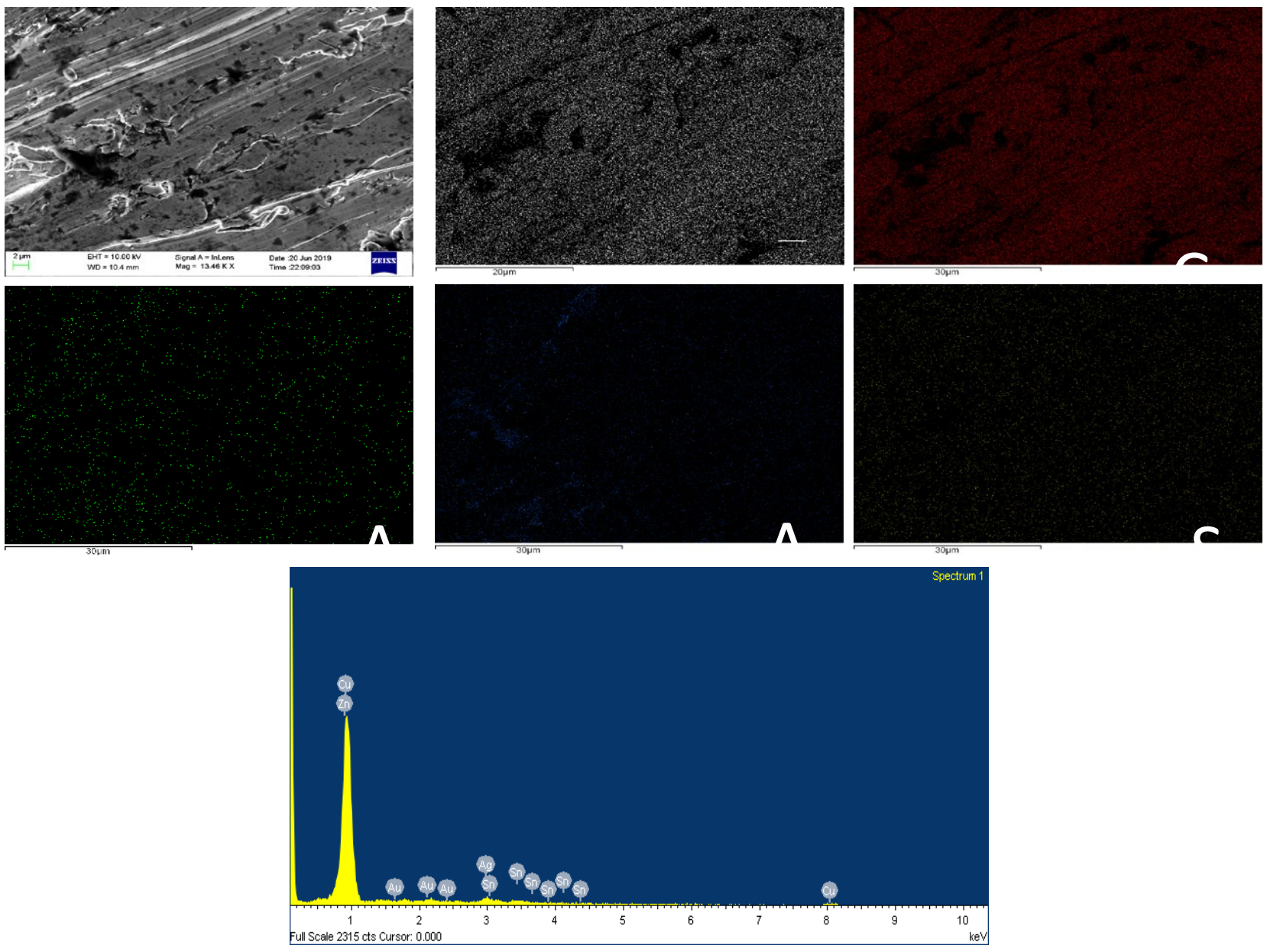

Figure 2: SEM mapping of Cu, Zn, Sn, Au and Ag with EDX analysis.

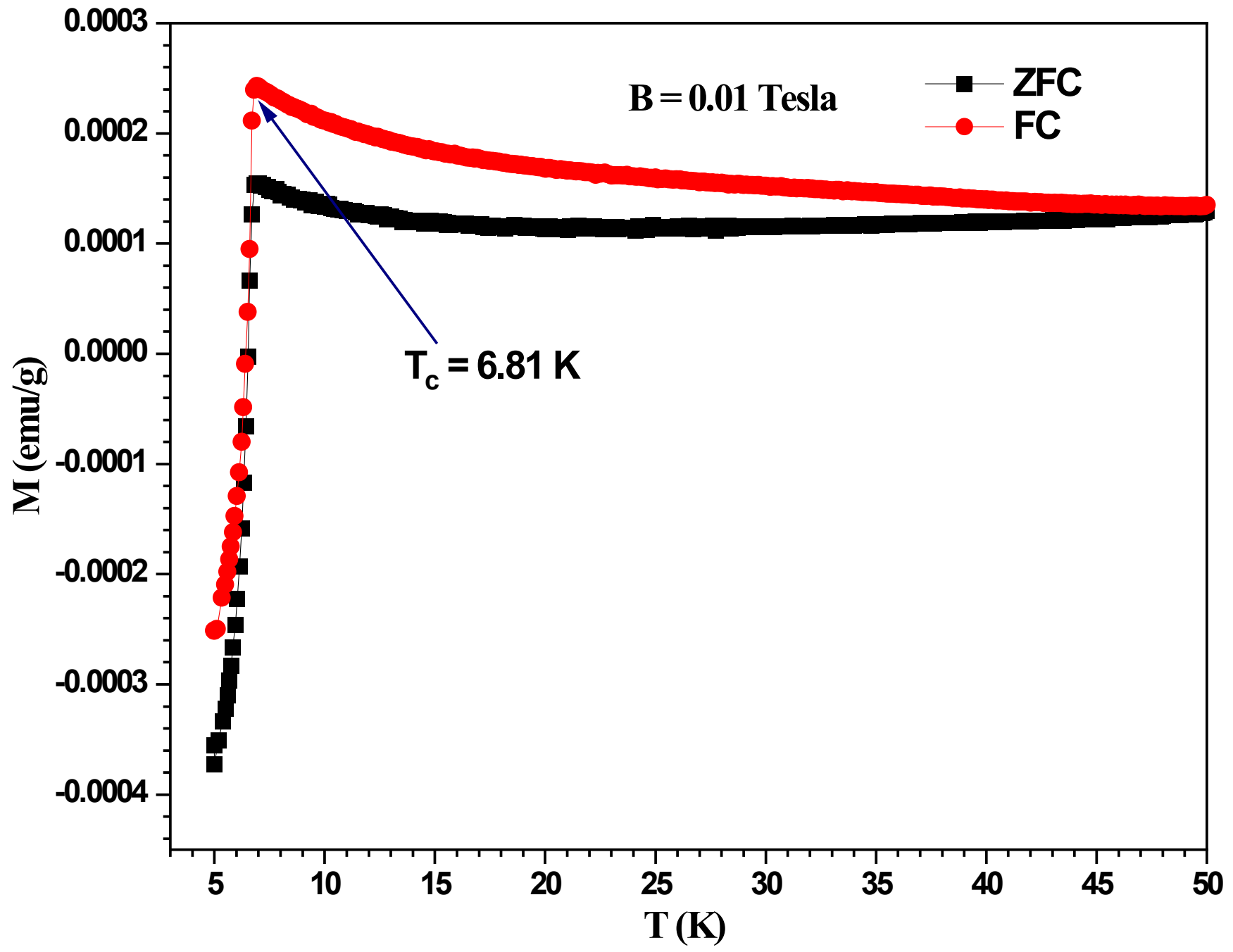

Figure 3: Magnetization (M) versus temperature (T) curves measured under Zero-Field-Cooled (ZFC) and Field-Cooled (FC) conditions at an applied magnetic field of . The sharp transition points to a critical temperature $T_c$=6.81 K.

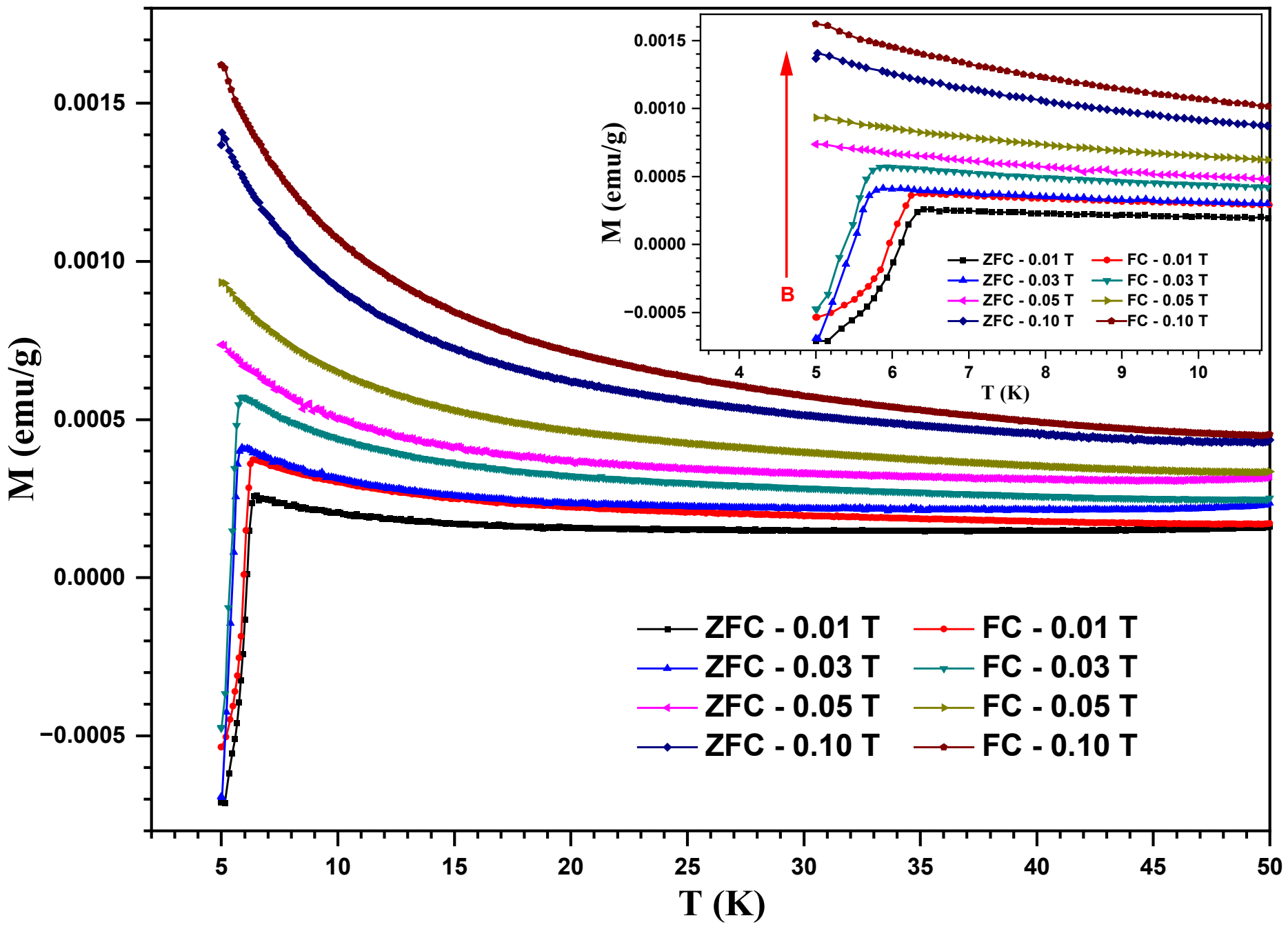

Figure 4: Temperature dependence of the magnetization (M) for various applied magnetic fields, , ranging from to under both ZFC and FC protocols. The data illustrate the suppression of the transition temperature with increasing field. The inset provides a magnified view of the transition region.

Figure 3 shows the temperature-dependent magnetization data measured under zero-field-cooled (ZFC) and field-cooled (FC) conditions at an applied magnetic field of 0.01 T. A distinct superconducting transition is observed in the ZFC curve, marked by the onset of strong diamagnetism below 7 K. The critical temperature ($T_c$), defined at the point of maximum slope in the ZFC curve, is estimated to be $T_c = 6.81$ K. Figure 4 displays the evolution of the magnetic response with increasing applied magnetic field from 0.01 to 0.1 T. Both the superconducting transition temperature and the magnitude of the diamagnetic signal decrease systematically with field, confirming the Type-II nature of the superconducting phase. A clear bifurcation between ZFC and FC curves persists below $T_c$ at all applied fields, indicating magnetic flux pinning within the mixed state. The isothermal magnetization curve *M(H)* recorded at 5 K (Figure 5) exhibits a pronounced hysteresis loop, characteristic of a Type-II superconductor with strong vortex pinning. The inset of Fig. 5 provides an enlarged view of the low-field region, showing the deviation from the initial linear (Meissner) response. The lower critical field ($H_{c1}$), defined as the field where this deviation occurs, is estimated to be approximately 0.018 T at 5 K. The hysteretic behaviour observed at low fields further supports the presence of trapped magnetic flux, consistent with the ZFC/FC separation seen in the magnetization data.

Figure 6 shows the temperature dependence of the electrical resistivity, ρ(T), measured between 300 and 2 K. The alloy exhibits metallic behaviour across the entire temperature range, with resistivity decreasing monotonically upon cooling. No abrupt drop to zero resistivity is observed down to 2 K, indicating that the primary conduction pathway in the bulk does not undergo a superconducting transition above this temperature. This suggests that any superconducting regions, if present, are limited in volume or spatially isolated, preventing the formation of a continuous superconducting path for current flow.

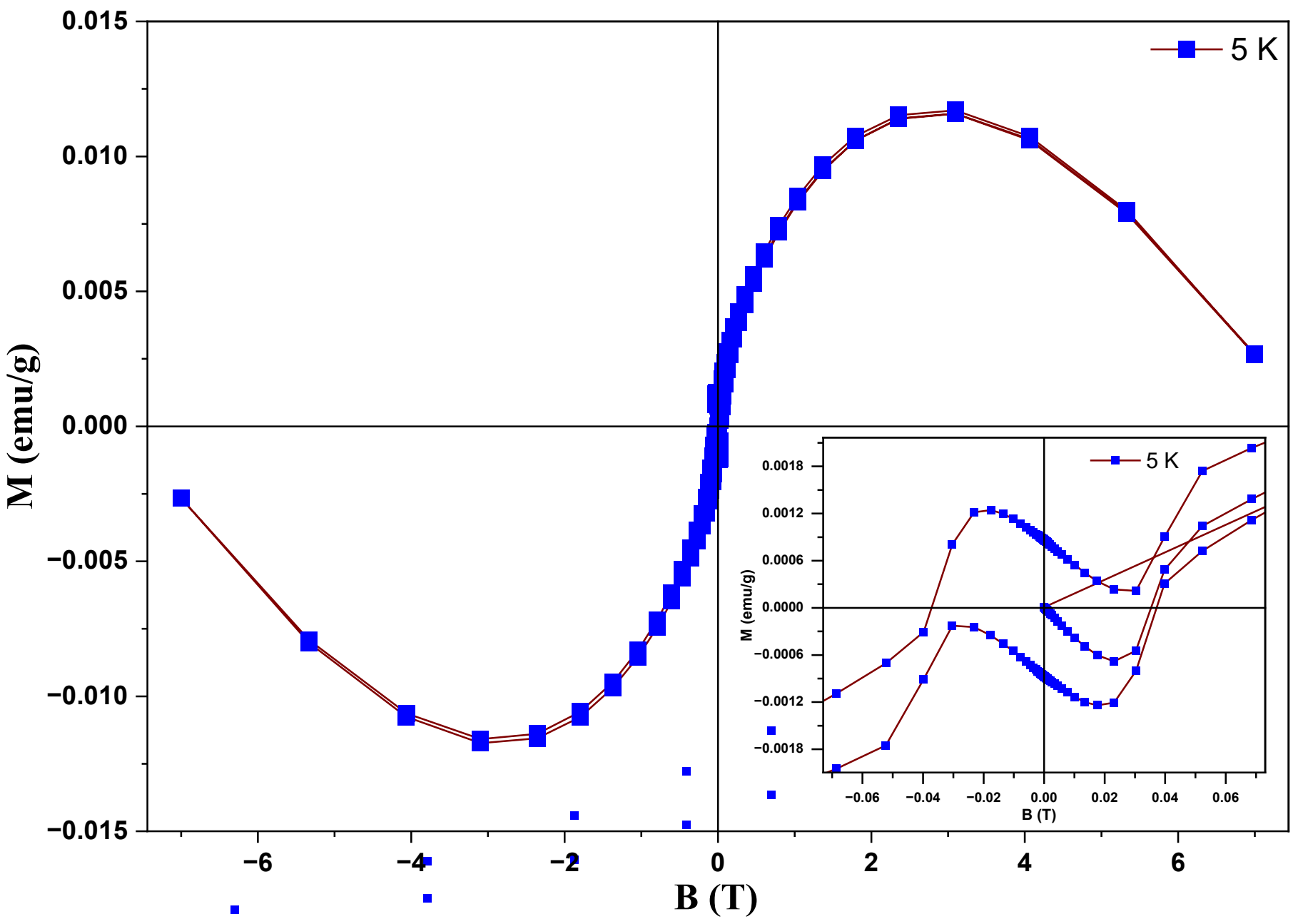

Figure 5: **Isothermal Magnetization (M) as a function of applied magnetic field (B)** measured at a temperature of . The data exhibit a hysteresis loop, characteristic of magnetic ordering. The inset shows a magnified view of the low-field region, highlighting the coercive field and remnant magnetization.

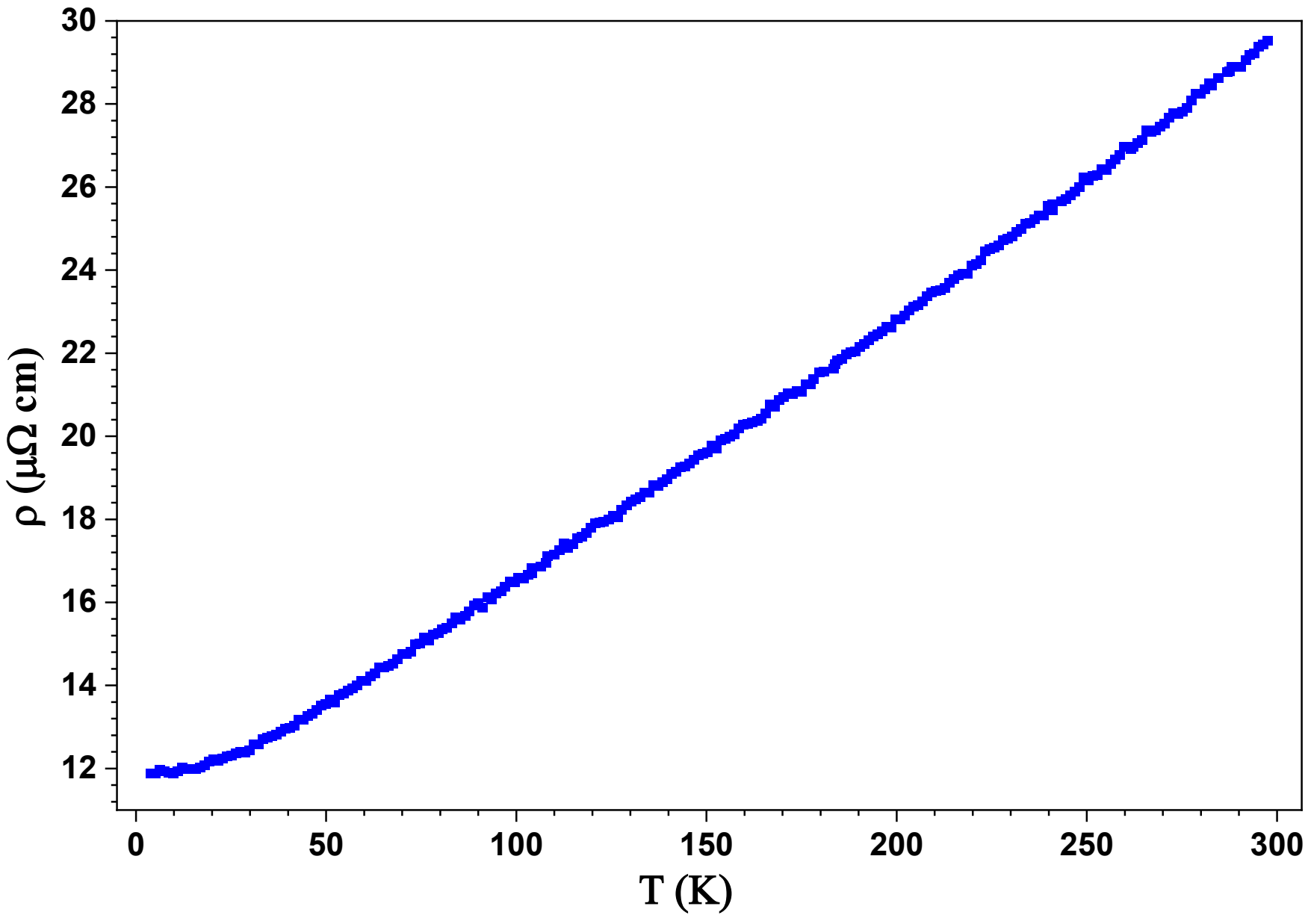

Figure 6: Temperature dependence of the electrical resistivity, , for the sample from 300 K to 2 K.

The coexistence of a clear diamagnetic onset in the magnetization data and the absence of a resistive transition is a hallmark of non-percolative superconductivity. The superconducting regions are likely confined to localized domains or thin filaments within the predominantly metallic Cu-rich matrix. The strong negative $M_{ZFC}$ signal confirms the presence of intrinsic superconductivity, whereas the finite residual resistivity demonstrates that the superconducting domains do not connect macroscopically. This behaviour likely originates from nanoscale compositional fluctuations or local strain fields observed in the FESEM/EDX analysis (Fig.2). Such local structural environments may favour electron pairing, forming isolated superconducting pockets within the disordered solid solution. The observation of a superconducting signal in a highly Cu-rich system is particularly intriguing, as elemental copper itself is non-superconducting. These results imply that the synergistic combination of Zn, Sn, Ag, and Au with Cu can create electronically favourable local environments capable of supporting superconductivity, even without long-range coherence.

## 4. Conclusion:

In summary, a Cu-rich multielement alloy with nominal composition Cu ≈ 80 wt%, Zn ≈ 20 wt%, Sn ≈ 4 wt%, Ag ≈ 8 wt%, and Au ≈ 2 wt% was synthesized using vacuum-sealed melt-quenching. Structural characterization confirmed the formation of a single-phase *fcc* Cu-type structure (*Fm–3m*) with a lattice parameter of $a$ = 3.6955 Å. Microstructural analysis revealed a fine-grained morphology with a near-uniform elemental distribution, though local compositional variations were evident. Electrical resistivity measurements confirmed metallic behaviour down to 2 K with no zero-resistance transition, whereas magnetization data exhibited a clear diamagnetic onset at $T_c$ = 6.81 K, characteristic of superconductivity. The contrasting magnetic and transport responses indicate that superconductivity in this system is confined to non-percolative, localized domains within the Cu-rich matrix. The estimated lower critical field ($H_{c1}$ ≈ 0.018 T) and magnetic hysteresis confirm Type-II superconductivity with strong flux pinning. This finding represents, to the best of our knowledge, the first observation of intrinsic superconductivity in a highly Cu-rich multielement alloy. The results highlight that alloying Cu with elements such as Zn, Sn, Ag, and Au can induce localized superconducting behaviour by altering the local electronic structure and lattice environment.

Future studies will focus on identifying the microscopic origin of this superconductivity using local structural probes such as X-ray absorption spectroscopy (XAS) and transmission electron microscopy (TEM), as well as investigating the pressure and compositional dependence of the superconducting phase fraction to understand its intrinsic mechanism.

**Acknowledgements**

AS wishes to thank DST-SERB for financial support. The author thank Dr. Boby Joseph of the Xpress beamline, Elettra Sincrotrone Trieste, for several fruitful discussions and XRD data collection.